\documentclass[manuscript]{emulateapj}
\usepackage{apjfonts}
\usepackage{natbib}

\shorttitle{X-ray limits on SN~2011fe}
\shortauthors{Margutti et al.}

\begin{document}
\title{Inverse Compton X-ray Emission from Supernovae with Compact Progenitors: Application to SN2011fe}
\author{R. Margutti\altaffilmark{1}, A.~M. Soderberg\altaffilmark{1}, L.~Chomiuk\altaffilmark{1,2}, R.~Chevalier\altaffilmark{3}, K.~Hurley\altaffilmark{4}, D.~Milisavljevic\altaffilmark{1}, R.~J. Foley\altaffilmark{1,21}, J.~P. Hughes\altaffilmark{5}, P.~Slane\altaffilmark{1}, C.~Fransson\altaffilmark{6}, M.~Moe\altaffilmark{1}, S. Barthelmy\altaffilmark{7}, W. Boynton\altaffilmark{8}, M. Briggs\altaffilmark{9},  V. Connaughton\altaffilmark{9}, E. Costa\altaffilmark{10}, J. Cummings\altaffilmark{7}, E. Del Monte\altaffilmark{10}, H. Enos\altaffilmark{8}, C. Fellows\altaffilmark{8}, M. Feroci\altaffilmark{10}, Y. Fukazawa\altaffilmark{11}, N. Gehrels\altaffilmark{7}, J. Goldsten\altaffilmark{12}, D. Golovin\altaffilmark{13}, Y. Hanabata\altaffilmark{11}, K. Harshman\altaffilmark{8}, H. Krimm\altaffilmark{7}, M. L. Litvak\altaffilmark{13}, K. Makishima\altaffilmark{14}, M. Marisaldi\altaffilmark{15},  I. G. Mitrofanov\altaffilmark{13}, T. Murakami\altaffilmark{13}, M. Ohno\altaffilmark{11}, D.~M.  Palmer\altaffilmark{17}, A. B. Sanin\altaffilmark{13}, R. Starr\altaffilmark{7}, D. Svinkin\altaffilmark{18}, 
T. Takahashi\altaffilmark{11}, M. Tashiro\altaffilmark{19}, Y. Terada\altaffilmark{19}, K. Yamaoka\altaffilmark{20}}

\altaffiltext{1}{Harvard-Smithsonian Center for Astrophysics, 60 Garden St., Cambridge, MA 02138, USA.}
\altaffiltext{2}{National Radio Astronomy Observatory, P. O. Box O Socorro, NM 87801, USA.}
\altaffiltext{3}{Department of Astronomy, University of Virginia, Charlottesville, VA 22904-4325, USA.}
\altaffiltext{4}{Space Sciences Laboratory, University of California, 7 Gauss Way, Berkeley, CA 94720-7450, USA.}
\altaffiltext{5}{Department of Physics and Astronomy, Rutgers University, Piscataway, NJ 08854-8019, USA.} 
\altaffiltext{6}{Department of Astronomy, Stockholm University, AlbaNova, SE-106 91 Stockholm, Sweden.}
\altaffiltext{7}{NASA/Goddard Space Flight Center Greenbelt, MD 20771, USA.}
\altaffiltext{8}{Department of Planetary Sciences, University of Arizona, Tucson, AZ 85721, USA.}
\altaffiltext{9}{Physics Department, The University of Alabama in Huntsville, Huntsville, AL 35809, USA.}
\altaffiltext{10}{INAF/IASF-Roma, via Fosso del Cavaliere 100, 00133 Roma, Italy.}
\altaffiltext{11}{Department of Physics, Hiroshima University, 1-3-1 Kagamiyama, Higashi-Hiroshima, Hiroshima 739-8526, Japan.}
\altaffiltext{12}{Applied Physics Laboratory, Johns Hopkins University, Laurel, MD 20723, USA.}
\altaffiltext{13}{Space Research Institute, 84/32, Profsoyuznaya, Moscow 117997, Russian Federation.}
\altaffiltext{14}{Department of Physics, University of Tokyo, 7-3-1 Hongo, Bunkyo-ku, Tokyo 113-0033, Japan.}
\altaffiltext{15}{INAF/IASF-Bologna, Via Gobetti 101, I-40129 Bologna, Italy.}
\altaffiltext{16}{Department of Physics, Kanazawa University, Kadoma-cho, Kanazawa, Ishikawa 920-1192, Japan.}
\altaffiltext{17}{Los Alamos National Laboratory, P.O. Box 1663, Los Alamos, NM 87545, USA.}
\altaffiltext{18}{Ioffe Physical-Technical Institute of the Russian Academy of Sciences, St. Petersburg, 194021, Russia.}
\altaffiltext{19}{Department of Physics, Saitama University, 255 Shimo-Okubo, Sakura-ku, Saitama-shi, Saitama 338-8570, Japan.}
\altaffiltext{20}{Department of Physics and Mathematics, Aoyama Gakuin University, 5-10-1 Fuchinobe, Sagamihara, Kanagawa 229-8558, Japan.}
\altaffiltext{21}{Clay Fellow.}
\date{Accepted YEAR month day. Received YEAR month day; in original form YEAR month day}

\begin{abstract}
 We present a generalized analytic
  formalism for the inverse Compton X-ray emission from hydrogen-poor
  supernovae and
  apply this framework to SN\,2011fe using Swift-XRT, UVOT
  and Chandra observations. We characterize
  the optical properties of SN\,2011fe in the Swift bands and find
  them to be broadly consistent with a ``normal'' SN Ia, however, no
  X-ray source is detected by either  XRT or Chandra.
  We constrain the progenitor system mass loss rate  $\dot M< 2\times
  10^{-9}\rm{M_{\sun}yr^{-1}}$ ($3\,\sigma$ c.l.) for wind velocity
  $v_w=100\,\rm{km~s^{-1}}$.  
  Our result rules out symbiotic binary progenitors for
  SN~2011fe and argues against Roche-lobe overflowing subgiants and
  main sequence secondary stars \emph{if} $\gtrsim 1\%$ of the
  transferred mass is lost at the Lagrangian points.  Regardless of
  the density profile, the X-ray non-detections are suggestive of a clean
  environment ($n_{CSM} < 150\, \rm{cm^{-3}}$) for $2\times
  10^{15}\lesssim R\lesssim 5\times 10^{16}$ cm around the progenitor
  site.  This is either consistent with the bulk of material
  being confined within the binary system or
  with a significant delay between mass loss and supernova
  explosion. We furthermore combine X-ray and radio limits from 
   Chomiuk et al. 2012 to constrain the post shock energy
 density in magnetic fields. Finally, we searched for the shock breakout pulse using
  gamma-ray observations from the Interplanetary Network and find no
  compelling evidence for a supernova-associated burst. Based on the
  compact radius of the progenitor star 
  we estimate that the shock break out pulse was likely not detectable
  by current satellites.  
\end{abstract}

\keywords{radiation mechanisms: non thermal}
\section{Introduction}
\label{Sec:Intro}

Over the past two decades,  the utility of Type Ia supernovae (SNe Ia) as standardizable candles
to trace the expansion history of the Universe has been underscored by the increasing resources 
dedicated to optical/near-IR discovery and follow-up campaigns (\citealt{Riess98}; \citealt{Perlmutter99}).  
 At the same time,  the nature of their 
progenitor system(s)  has remained elusive, 
despite aggressive studies to unveil them (see e.g. \citealt{Hillebrandt00}).  
The second nearest Ia SN discovered in the digital era,
SN~2011fe \citep{Nugent11} located at $d_{\rm{L}}=6.4\,\rm{Mpc}$ \citep{Shappee11}, represents a 
natural test bed for a detailed SN Ia progenitor study\footnote{The nearest Type Ia in the digital era is SN~1986G which
exploded in NGC~5128 at a distance of $\sim4\,\rm{Mpc}$ \citep{Frogel87}.}. The best studied Type Ia SN at early times
before SN~2011fe, SN~2009ig, demonstrated how single events can provide significant insight into 
the properties of this class of explosions \citep{Foley12}.

The fundamental component of SN Ia progenitor models is an accreting 
white dwarf (WD) in a binary system.
Currently, the most popular models include 
 (i) a single-degenerate (hereafter, SD) scenario in which a massive WD accretes
material from a H-rich or He-rich companion, potentially a giant, subgiant or main-sequence star, 
(\citealt{Whelan73}; \citealt{Nomoto80}). Mass is transferred either via Roche-lobe overflow 
(RLOF) or through stellar winds.   Alternatively, (ii) models invoke a double sub-$M_{\rm Ch}$ WD binary system 
that eventually merges (double degenerate model, DD; \citealt{Iben84}, \citealt{Webbink84}). 

In SD models, the circumbinary environment may be enriched by the stellar wind of the donor 
star or through non-conservative mass transfer
in which a small amount of material is lost to the surroundings.
Winds from the donor star shape the local density profile as $\rho_{\rm CSM}\propto R^{-2}$ over a $\lesssim 1$ parsec  region 
encompassing the binary system. Theoretical considerations indicate that the wind-driven mass loss rate must be low, since an
accretion rate of just $\sim 3\times10^{-7}~\rm M_{\rm{\odot}}~\rm{yr^{-1}}$
is ideal for the WD to grow slowly up to $M_{\rm Ch}$ and still avoid
mass-losing nova eruptions (steady burning regime, \citealt{Nomoto84}). Strong evidence for the {\it lack} of a 
wind-stratified medium and/or the detection of a constant
local density (with a typical interstellar medium density of
$n_{\rm CSM}\approx 0.1-1~\rm cm^{-3}$) may instead point to a DD model.

Arising from the  interaction of the SN shock blast wave with the circumbinary  
material, radio and X-ray observations can potentially discriminate between
the two scenarios by shedding light on the properties of the environment, shaped by the evolution of the progenitor system
(see e.g. \citealt{Boffi95}, \citealt{Eck95}).  
Motivated thus, several dozen SNe Ia at distances $d\lesssim 200$ Mpc 
have been observed with the Very Large Array (VLA; \citealt{Panagia06}; 
\citealt{Hancock11}; Soderberg in prep.),  the Chandra X-ray Observatory 
\citep{Hughes07}, and
the Swift X-ray Telescope (\citealt{Immler06}; Russel \& Immler, in press) revealing no detections to
date\footnote{We note that the claimed detection of SN\,2005ke with the \emph{Swift}-XRT was 
not confirmed with follow-up Chandra observations, strongly suggesting that the Swift/XRT source 
was due to contamination from the host galaxy \citep{Hughes07}.}.   These limits were used to constrain the density 
of the circumbinary material, and in turn the mass loss rate of the progenitor system.  
However these  data poorly constrain the WD companion, due in part to the limited sensitivity of the observations
(and the distance of the SNe). 
The improved sensitivity of the Expanded Very Large Array (EVLA) coupled with a more detailed approach regarding the relevant radio and X-ray emission (and absorption) processes in Type Ia supernovae,  has enabled the deepest constraints to date on a circumbinary progenitor as discussed in our companion paper on the recent Type Ia SN\,2011fe/ PTF11kly (\citealt{Chomiuk12}. See also \citealt{Horesh11}).

Here we report a detailed panchromatic study of SN\,2011fe bridging
optical/UV and gamma-ray observations.   Drawing from observations
with the \emph{Swift} and Chandra satellites as well as the
Interplanetary Network (IPN; \citealt{Hurley10}), we constrain the
properties of the bulk ejecta and circumbinary environment through a 
self-consistent characterization of the dynamical evolution of the shockwave.
First we present optical/UV light-curves for the SN, indicating that
the object appears consistent with a "normal" SN Ia.  Next we
discuss deep limits on the X-ray emission in the month following
explosion.  We furthermore report gamma-ray limits (25-150 keV) for
the shock breakout pulse.  In the Appendix we present an analytic
generalization for the the Inverse Compton (IC) X-ray luminosity
expected from hydrogen poor SNe that builds upon previous work by
\citealt{Chevalier06} and \citealt{Chevalier06b}  but is broadly
applicable for a wide range of shock properties, metallicity, photon
temperatures, and circumstellar density profiles (stellar wind or ISM;
see Appendix \ref{Sec:IClum}).    We apply this analytic model to
SN\,2011fe to constrain the density of the circumbinary environment,
and find that our limits are a factor of $\sim $ 10 deeper than the 
results recently reported by \citealt{Horesh11}.

Observations are described in Sec. \ref{Sec:Obs}; limits to the SN progenitor
system  from X-ray observations are derived and discussed in Sec. \ref{Sec:Xray}
using the IC formalism from Appendix \ref{Sec:IClum}. We combine our radio
\citep{Chomiuk12} and X-ray limits to constrain the post-shock energy density
in magnetic fields in Sec. \ref{Sec:epsilonB}, while the results from the search 
of a burst of gamma-ray radiation from the SN shock break-out is presented in
Sec. \ref{Sec:Gammaray}. Conclusions are drawn in Sec. \ref{Sec:conc}.
\section{Observations}
\label{Sec:Obs}

\begin{figure}
\vskip -0.0 true cm
\centering
    \includegraphics[scale=0.6]{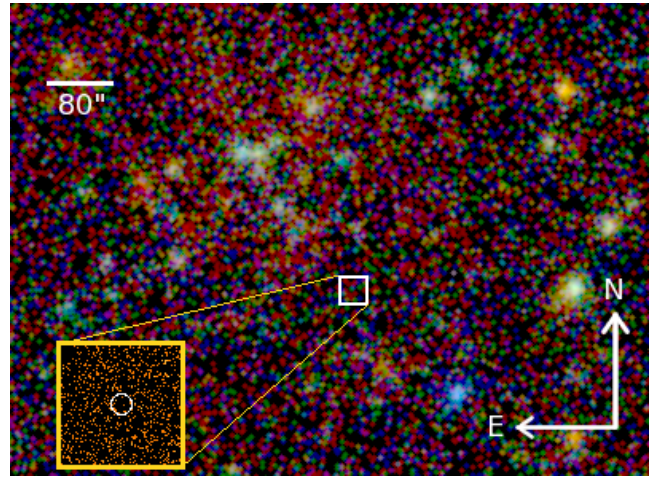}
     \caption{\emph{Swift}-XRT color combined image of the environment around SN~2011fe. 
     Red, green and
     blue colors refer to soft (0.3-1 keV), medium (1-3 keV) and hard (3-10 keV) 
     sources, respectively. A $40"$ region around the SN is marked with a white box.
     \emph{Inset:} \emph{Chandra} 0.5-8 keV deep observation of the same region obtained at
     day 4 since the explosion. No source is detected at the SN position (white circle). }
\label{Fig:X-rays}
\end{figure}

SN~2011fe was discovered by the Palomar Transient Factory (PTF) on 2011 August 24.167 UT 
and soon identified as a very young type Ia explosion in the Pinwheel galaxy (M101) (\citealt{Nugent11b}).
From early time optical observations \cite{Nugent11} were able to constrain the SN explosion date 
to August 23, $16:29\pm20\,\rm{min}$ (UT).
The SN site was fortuitously observed both by the \emph{Hubble Space Telescope} (HST) and by \emph{Chandra}
on several occasions prior to the explosion in the optical and X-ray band, giving the possibility 
to constrain the progenitor system (\citealt{Li11}; \citealt{Liu11}). Very early optical and UV 
photometry has been used by \cite{Brown11} and \cite{Bloom11} to infer  the progenitor and 
companion radius and nature, while multi-epoch high-resolution spectroscopy taken during the evolution of the
SN has been employed as a probe of the circumstellar environment \citep{Patat11b}. Limits to the
circumstellar density have been derived from deep radio observations in our companion paper \citep{Chomiuk12},
where we consistently treat the shock parameters and evolution. Here we study SN~2011fe
from a complementary perspective, bridging optical/UV, X-ray and gamma-ray observations. 
  
\emph{Swift} observations were acquired starting from August 24, $1.25$
days since the onset of the explosion.
\emph{Swift}-XRT data have been analyzed using the latest release of the HEASOFT package at the
time of writing (v11). Standard filtering and screening criteria have been applied.
No X-ray source consistent with the SN position 
is detected in the 0.3-10 keV band either in promptly available data (\citealt{Horesh11}; \citealt{Margutti11a})
 or in the combined $142$ ks 
exposure covering the time interval $1-65$ days (see Fig. \ref{Fig:X-rays}).  In particular, using the first 4.5 ks obtained on
August 24th, we find a PSF (Point Spread Function) and exposure map corrected\footnote{Note that
correcting for both the PSF \emph{and} the exposure map is here of primary importance to compute
the upper limits. If the exposure map is neglected, deeper but unrealistic limits would be computed.}  
$3\sigma$ count-rate limit on the undetected SN $\lesssim4\times 10^{-3}\,\rm{c\,s^{-1}}$. 
For a simple power-law spectrum with photon index $\Gamma \sim 2$ and Galactic
neutral hydrogen column density $\rm{NH}=1.8\times 10^{20}\,\rm{cm^{-2}}$ \citep{Kalberla05}
this translates into an unabsorbed 0.3-10 keV flux $F=1.5\times 10^{-13}\,\rm{erg\,s^{-1}cm^{-2}}$
corresponding to a luminosity  $L=7\times 10^{38}\,\rm{erg\,s^{-1}}$  at a 
distance of 6.4 Mpc \citep{Shappee11}.  Collecting data between 1 and 65 days after the explosion
(total exposure of $142$ ks) we obtain a $3\sigma$ upper limit of $2\times 10^{-4}\,\rm{c\,s^{-1}}$ 
($F=7.4\times 10^{-15}\,\rm{erg\,s^{-1}cm^{-2}}$, $L=3.6\times 10^{37}\,\rm{erg\,s^{-1}}$).
Finally, extracting data around maximum light  (the time interval  8-38 days),
the X-rays are found to contribute less than 
$3\times 10^{-4}\,\rm{c\,s^{-1}}$  ($3\sigma$ limit, total exposure of 61 ks) corresponding to
$F=1.1\times 10^{-14}\,\rm{erg\,s^{-1}cm^{-2}}$, $L=5.9\times 10^{37}\,\rm{erg\,s^{-1}}$.

\begin{figure*}
\vskip -0.0 true cm
\centering
    \includegraphics[scale=0.7]{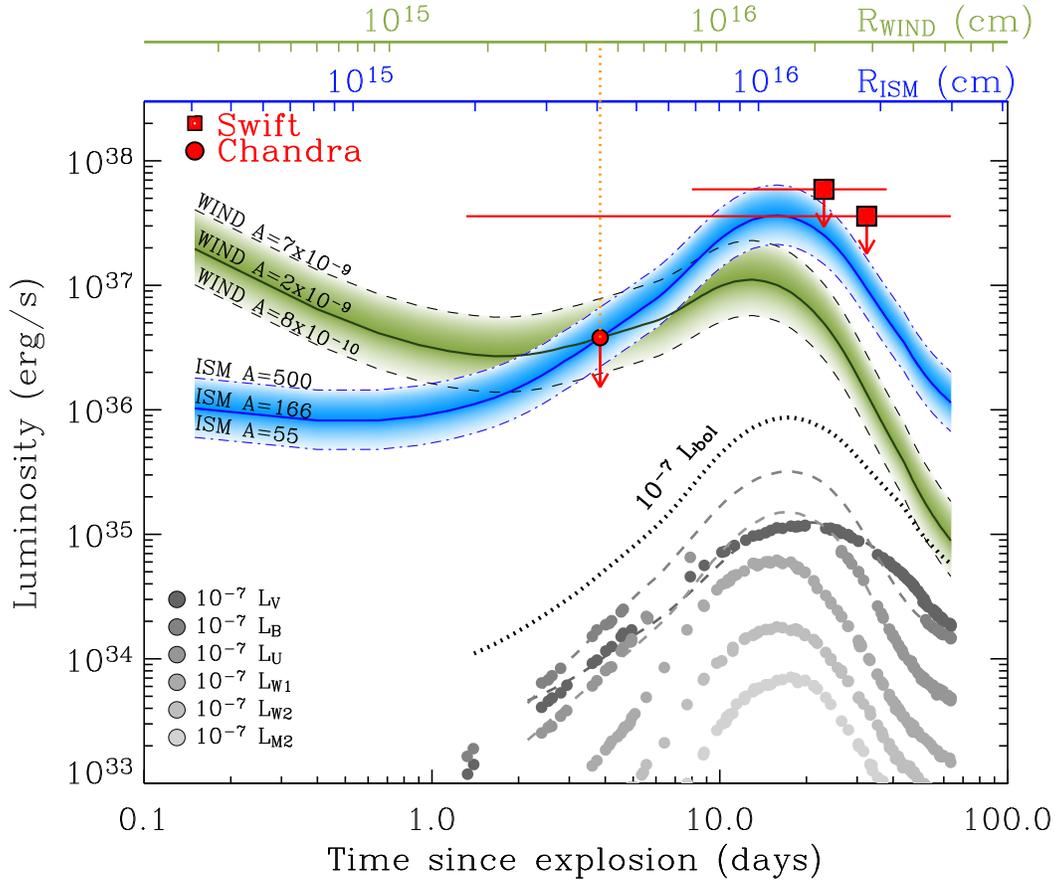}
     \caption{Limits on the X-ray luminosity of SN~2011fe: 0.5-8 keV luminosity expected from 
     inverse comptonization of optical photons  in the case of
     a wind $\rho_{CSM} \propto R^{-2}$ (green solid line) and an ISM $\rho_{CMS}\propto const$  (blue solid
     line) environment. Deep limits from \emph{Swift} and \emph{Chandra} are marked with 
     red bullets and squares, respectively.  In the case of \emph{Swift} observations we report the 
     combined limit (at the linear midpoint of the time intervals), 
     produced stacking the entire \emph{Swift}-XRT data set together with a limit
     calculated around the SN maximum light. The colored areas span 
     $A=(0.8-7)\times10^{-9}\,M_\odot yr^{-1}/(100\,\rm{km\,s^{-1}})$ (wind, green) and
     $A=(55-500)\, \rm{cm^{-3}}$ (ISM, blue).
     The \emph{Chandra} observation constrains  $\dot M/v_w<2\times 10^{-9}\,M_\odot yr^{-1}/(100\,\rm{km\,s^{-1}})$ (wind);
      $n_{CSM}<166\,\rm{cm^{-3}}$ (ISM). The blue and green x-axes report the ISM and wind
      radius of the shock calculated using these values.
       Black dotted line: scaled SN bolometric luminosity. 
       Grey filled circles: scaled \emph{Swift}-UVOT light-curves. Dashed lines: best-fitting
       \cite{Nugent02} templates to the u b and v band. 
       We assume $E=10^{51}\,\rm{erg}$, $M_{ej}=1.4\,M_{\sun}$, $\epsilon_e=0.1$, $p=3$.}
\label{Fig:IC}
\end{figure*}

We observed SN~2011fe with the \emph{Chandra} X-ray Observatory on Aug 27.44 UT (day 4 since
the explosion) under an approved DDT proposal (PI Hughes).
Data have been reduced with the CIAO software package (version 4.3), with calibration
database CALDB (version 4.4.2).  We applied standard filtering using
CIAO threads for ACIS data.
No X-ray source is detected at the SN position during the 50 ks exposure \citep{Hughes11}, with a $3\sigma$
upper limit of  $1.1\times 10^{-4}\,\rm{c\,s^{-1}}$ in the 0.5-8 keV band, from which we derive 
a flux limit of  $7.7\times 10^{-16}\,\rm{erg\,s^{-1}cm^{-2}}$ corresponding to 
$L=3.8\times 10^{36}\,\rm{erg\,s^{-1}}$ (assuming a simple power-law model with
spectral photon index $\Gamma=2$). $3\sigma$ upper limits from \emph{Swift} and 
\emph{Chandra} observations are shown in Fig.  \ref{Fig:IC}.

The SN was  clearly detected in \emph{Swift}-UVOT observations.
Photometry was extracted from a $5\arcsec$ aperture, closely following
the prescriptions by \cite{Brown09} (see Fig. \ref{Fig:IC}). Pre-explosion images of the host galaxy acquired by 
UVOT in 2007 were used to estimate and subtract the host galaxy light contribution. 
Our photometry agrees (within the uncertainties) with the results of \cite{Brown11}. 
With respect to \cite{Brown11} we extend the UVOT photometry of SN~2011fe
to day $\sim60$ since the explosion.
Due to the brightness of SN~2011fe, u, b and v observations strongly suffer from coincidence
losses \citep{Breeveld10} around maximum light (see \citealt{Brown11} for details): 
supernova templates from \cite{Nugent02} were used to fit the u and b light-curves and infer the SN luminosity 
during those time intervals in the u and b bands. For the v-band, it was possible to (partially) recover the
original light-curve applying standard coincidence losses corrections: however, due to
the extreme coincidence losses, our v-band light-curve may still provide a lower limit to the real 
SN luminosity in the time interval $8-37$ days since explosion. In Fig. \ref{Fig:IC} we present the  
\emph{Swift}-UVOT  6-filter light-curves, and note that the re-constructed v-band is broadly 
consistent with the Nugent template\footnote{Note that, as it will be clear from the
next section, this possible underestimation of the 
v-band luminosity around maximum light only leads to \emph{more conservative} limits to the ambient
density derived from \emph{Swift} observations. Our main conclusions are however based on the \emph{Chandra}
observation taken at day 4, when coincidence losses do \emph{not} play a role.}. 
We adopted a Galactic reddening of $E(B-V)=0.01$ \citep{Schlegel98}.

In the case of the "golden standard"  Ia SN~2005cf (which is among the best studied Ia SNe), 
the V band is found to contribute $\sim20\%$ to the bolometric luminosity \citep{Wang09}, with limited variation
over time. For SN~2011fe, we measure at day 4 a v-band luminosity  $L_{v}\sim 10^{41}\,\rm{erg\,s^{-1}}$, corresponding to
$L_{\rm{bol}}\approx 5\times 10^{41}\,\rm{erg~s^{-1}}$ and note that at this time the luminosity in the v, b, u, w1 and w2 bands
account for $\approx 0.5\,L_{\rm{bol}}$. We therefore assumed that the v, b, u, w1 and w2 bands represent\footnote{Nearly 
80\% of the bolometric luminosity of a typical SN Ia is emitted in the range from 3000 to 10000 \AA  
\citep{Contardo00}.}
$\approx 0.5 L_{\rm{bol}}$.
In the following  we explicitly provide the dependence of our density limits on $L_{\rm{bol}}$, 
so that it is easily possible to re-scale our limits to any $L_{\rm{bol}}$ value.
Given that the optical properties point to a normal SN Ia
(Parrent at al. in prep.) we adopt fiducial parameters
$M_{ej}=1.4\,\rm{M_{\sun}}$ and $E=10^{51}\,\rm{erg}$ 
for the ejecta mass and SN energy, respectively, throughout this paper.

\section{Limits on the ambient density from X-rays}
\label{Sec:Xray}

\begin{figure}
\vskip -0.0 true cm
\centering
    \includegraphics[scale=0.45]{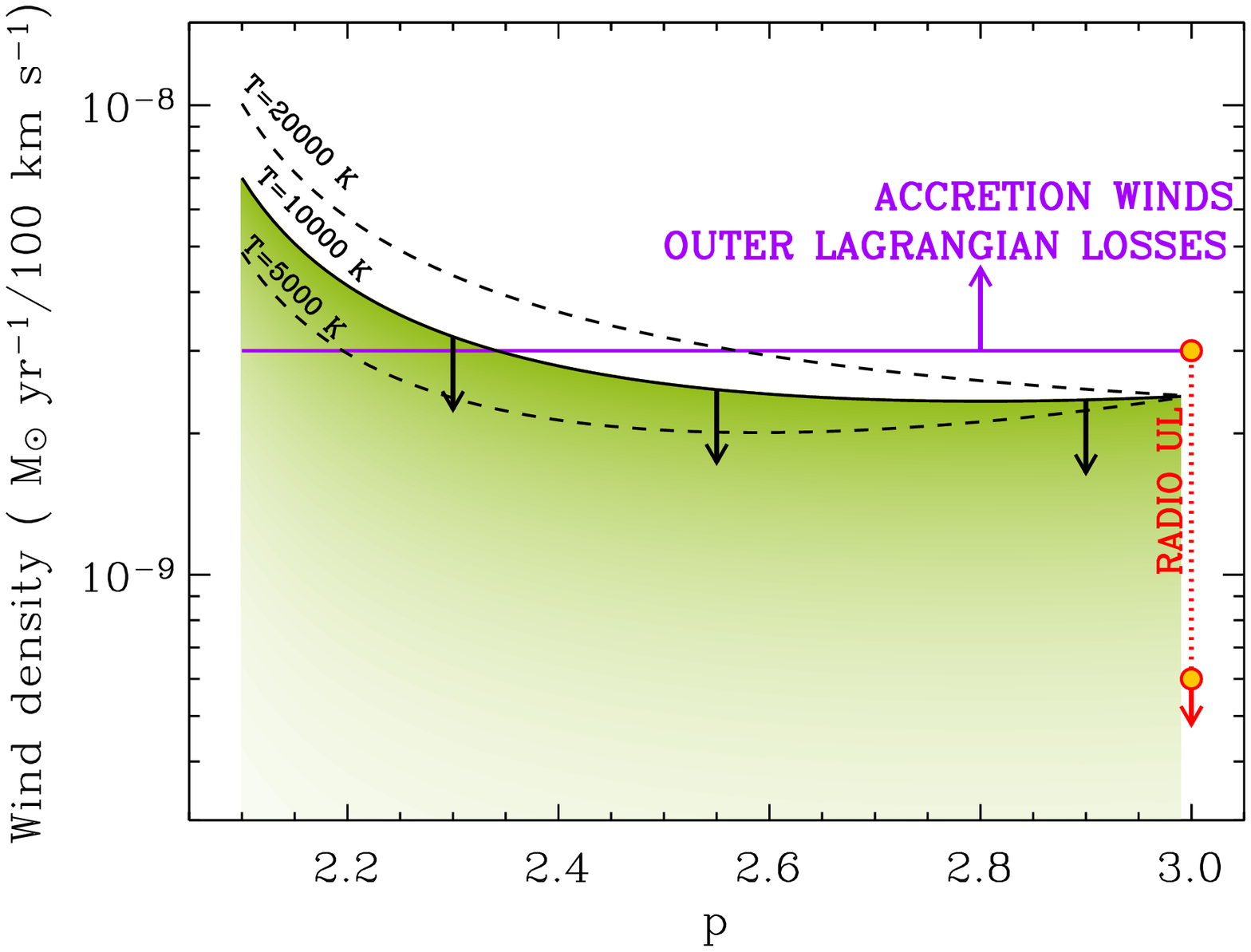} 
    \includegraphics[scale=0.45]{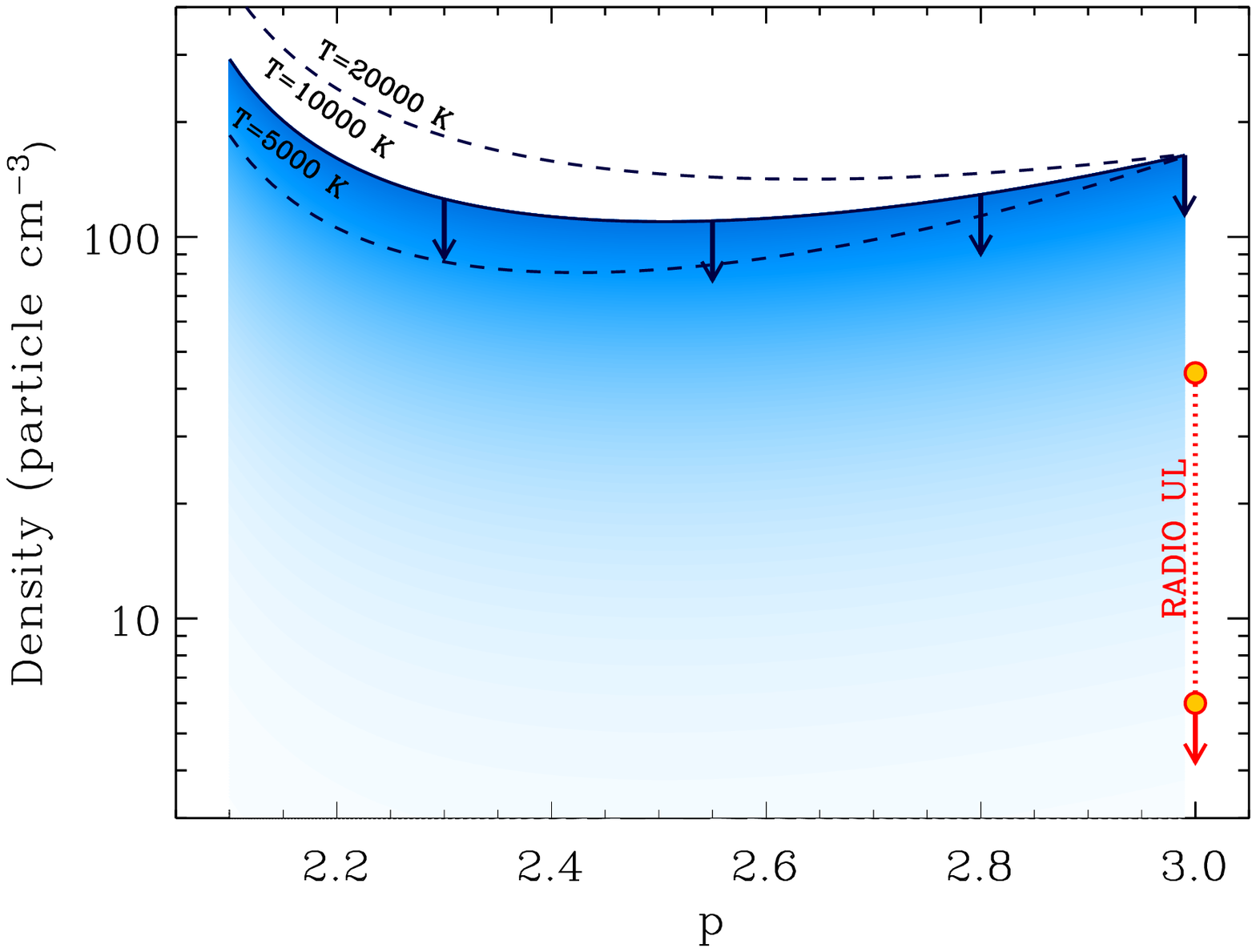}
     \caption{Limits on the CSM density around SN~2011fe as derived from the X-ray non-detection at 4 days after the 
     explosion, assuming inverse comptonization of optical photons in the case of a wind (upper panel)
     or ISM (lower panel) scenario. Black solid line: $3\,\sigma$ 
     upper limit as a function of the power-law index of the electron distribution $p$ assuming $T=10000\,\rm{K}$.
     Upper limit contours in the cases $T=5000$ K and
     $T=20000$ K are also shown for comparison (black dashed lines).   Yellow bullets: upper limit to the CSM
     density as derived from radio observations for $\epsilon_B$ in the range $0.1-0.01$. $\epsilon_B=0.1$ gives
     the tightest constraint \citep{Chomiuk12}.
     We assume $E=10^{51}\,\rm{erg}$, $M_{ej}=1.4\,M_{\sun}$, $\epsilon_e=0.1$.  }
\label{Fig:ICWIND}
\end{figure}

X-ray emission from SNe may be attributed to a number of emission
processes including (i) synchrotron, (ii) thermal, (iii) Inverse Compton (IC), 
or (iv) a long-lived central engine (see \citealt{Chevalier06} for a review).
It has been shown that the X-ray emission from stripped supernovae exploding into
low density environments is dominated by IC on a timescale of
weeks to a month since explosion, corresponding to the peak of the optical
emission (\citealt{Bjornsson04}, \citealt{Chevalier06}).  In
specific cases, this has been shown to be
largely correct (e.g., SN~2008D \citealt{Soderberg08}, SN~2011dh \citealt{Soderberg11}).

In this framework the X-ray emission is originated by up-scattering of 
optical photons from the SN photosphere by a population of relativistic electrons (e.g. \citealt{Bjornsson04}).
The IC X-ray luminosity depends on the density structure of the SN ejecta, the structure of the
circumstellar medium (CSM) and the details of the relativistic electron distribution responsible
for the up-scattering. Here we assume the SN outer density structure 
$\rho_{\rm{SN}}\propto R^{-n}$ with $n\sim10$ \citep{Chevalier06},
as found for SNe arising from compact progenitors (as a comparison, \citealt{Matzner99} found
the outermost profile of the ejecta to scale as $\rho_{\rm{SN}}\propto R^{-10.2}$. 
See \citealt{Chomiuk12}, Soderberg in prep. for a discussion)\footnote{Note that the adopted 
density profile is similar to the W7 model by \cite{Nomoto84} with the addition of a power-law profile
at high velocities. A pure W7 profile would give
rise to somewhat slower shockwave velocity \citep{Dwarkadas98}.};
the SN shock propagates into the circumstellar medium and is assumed to accelerate the 
electrons in a power-law  distribution 
$n_{e}(\gamma)=n_0\gamma^{-p}$ for $\gamma> \gamma_{\rm{min}}$. 
Radio observations of type Ib/c SNe indicate $p\sim3$ \citep{Chevalier06}. 
However, no radio detection has ever been obtained for a type Ia SN so that the value of $p$
is currently unconstrained: this motivates us to explore a wider parameter space 
$p\gtrsim 2.1$ (Fig. \ref{Fig:ICWIND}) as seen for mildly relativistic and relativistic explosions (e.g.,
gamma-ray bursts,  \citealt{Panaitescu00}; \citealt{Yost03}; \citealt{Curran10}).
Finally, differently from  the thermal or synchrotron mechanisms, the IC luminosity is directly related to
the bolometric luminosity of the SN ($L_{\rm{IC}}(t)\propto L_{\rm{bol}}(t)$): the environment
directly determines the \emph{ratio} of the optical to the X-ray luminosity, so that
possible uncertainties on the distance of the SN do not affect the IC computation;
it furthermore does \emph{not} require any assumption on magnetic field related parameters.

For a population of optical photons with effective temperature $T_{\rm{eff}}$, the IC
luminosity at frequency $\nu$ reads (see Appendix \ref{Sec:IClum}):
\begin{equation}
\label{Eq:ICgeneral}
	\frac{dL_{\rm{IC}}}{d\nu}\sim0.2\Big(\frac{h}{3.6k} \Big)^\frac{3-p}{2}\frac{(p-2)\sigma_T
	\epsilon_e\rho_{CMS} v_s^2 \gamma_{\rm{min}}^{(p-2)} T_{\rm{eff}}^{\frac{p-3}{2}} \nu^{\frac{1-p}{2}} \Delta R}{m_e c^2}L_{\rm{bol}}(t)
\end{equation}
where $\Delta R$ is the extension of the region containing fast electrons;
$\rho_{CSM}$ is the circumstellar medium density the SN shock is impacting on,
which we parametrize as a power-law in shock radius $\rho_{CSM}\propto R^{-s}$;
together with $\rho_{\rm{SN}}$, $\rho_{CSM}$  determines the shock dynamics, 
directly regulating the evolution of the shock velocity $v_{s}\equiv v_{s}(t,n,s)$,
shock radius $R\equiv R(t,n,s)$ and 
$\gamma_{\rm{min}}\equiv \gamma_{\rm{min}}(t,n,s)$ as derived in Appendix \ref{Sec:IClum}.  
For the special case $p=3$,
$\frac{dL_{\rm{IC}}}{d\nu}\propto \nu^{-1}$, its dependence on $T_{\rm{eff}}$
cancels out and it is straightforward to verify that Eq. \ref{Eq:ICgeneral} matches
the predictions from \cite{Chevalier06}, their Eq. (31) for $s=2$ (wind medium).
In the following we use Eq. \ref{Eq:ICgeneral} and the $L_{\rm{bol}}(t)$ evolution
calculated from \emph{Swift}-UVOT observations of SN~2011fe (Sec. \ref{Sec:Obs})
to derive limits on the SN environment assuming different density profiles. 
We assume $\epsilon_e=0.1$, as indicated by well studied SN
shocks \citep{Chevalier06}. Each limit on the environment density we report below
has to be re-scaled of a multiplicative factor $(0.1/\epsilon_e)^{(p-1)}$ for other 
 $\epsilon_e$ values.
\subsection{Wind scenario}
\label{SubSec:wind}

A star which has been losing material at constant rate $\dot M$ gives
rise to a "wind medium": $\rho_{CSM}=\dot M/(4\pi R^{2}v_w)$. 
Eq. \ref{Eq:ICWIND} and the \emph{Chandra} non-detection
constrain the wind density to 
$\dot M/v_w<2\times 10^{-9} (M_{\sun}y^{-1}/100\,\rm{km\,s^{-1}})$ 
(where $v_w$ is the wind velocity). This is a $3\sigma$ limit obtained integrating 
Eq. \ref{Eq:ICWIND} over the 0.5-8 keV \emph{Chandra} pass band and
assuming $p=3$, $\epsilon_e=0.1$, $E=10^{51}\,\rm{erg}$ and $M_{ej}=1.4\,M_{\sun}$.
The observation was performed on day 4 after the explosion:
at this time $L_{\rm{bol}}\sim5\times10^{41}\,\rm{erg\,\rm{s^{-1}}}$ while
the shock wave probes the environment density at a radius $R\sim4\times10^{15}\,\rm{cm}$
(Eq. \ref{Eq:vshockRshock} and \ref{Eq:vshockWIND}) for 
$\dot M/v_w=2\times 10^{-9} (M_{\sun}y^{-1}/100\,\rm{km\,s^{-1}})$ (see Fig. 
\ref{Fig:IC}). For the wind scenario $\dot M/v_w\propto (1/L_{\rm{bol}})^{(1/0.64)}$
(see Appendix \ref{Sec:IClum}). 

While giving less deep constraints, \emph{Swift} observations have the
advantage of being spread over a long time interval giving us the possibility to 
probe the CSM density over a wide range of radii. 
Integrating Eq.  \ref{Eq:ICWIND} in the time interval 1-65 days to 
match the \emph{Swift} coverage (and using the 0.3-10 keV band) leads
to $\dot M/v_w<7\times 10^{-8} (M_{\sun}y^{-1}/100\,\rm{km\,s^{-1}})$ for 
$2\times 10^{15}\lesssim R\lesssim 6\times 10^{16}\, \rm{cm}$ from the progenitor 
site\footnote{Given the gentle scaling of the shock radius
with wind density ($R\propto A^{-0.12}$, Eq. \ref{Eq:vshockWIND}), 
these values are accurate within a factor 10 of $\dot M/v_w$ variation.}.
A similar value is obtained using the X-ray limit around maximum optical light, 
when the X-ray emission from IC is also expected to peak (Fig. \ref{Fig:IC}\footnote{Note 
that in Fig. \ref{Fig:IC} the \emph{Swift} limits are arbitrarily assigned to the linear
midpoint of the temporal intervals. The limit on the ambient density is however 
calculated integrated the model over the entire time interval so that the arbitrary
assignment of the ``central'' bin time has no impact on our conclusions.}).

\subsection{ISM scenario}
\label{SubSec:ism}
SN~2011fe might have exploded in a uniform density environment (ISM, $s=0$). In this case,
integrating Eq. \ref{Eq:ICISM} over the 0.5-8 keV energy range, 
the \emph{Chandra} limit implies a CSM density $n_{CSM} <166\,\rm{cm^{-3}}$
at $3\sigma$ confidence level for fiducial parameter values $p=3$, $\epsilon_e=0.1$, 
$E=10^{51}\,\rm{erg}$ and $M_{ej}=1.4\,M_{\sun}$. This limit applies to day 4 after
the explosion (or, alternatively to a distance $R\sim4\times 10^{15}\,\rm{cm}$, see
Fig. \ref{Fig:IC}). Integrating Eq. \ref{Eq:ICISM} over the time interval 1-65 days
(and in the energy window 0.3-10 keV) the \emph{Swift} upper limit implies 
$n_{CSM}<800 \,\rm{cm^{-3}}$ ($3\sigma$ level), over a distance range 
$2\times10^{15}-3\times 10^{16}\,\rm{cm}$ from the progenitor site\footnote{$R$ 
has a very gentle ($\propto A^{-0.1}$, see Eq. \ref{Eq:vshockISM}) dependence on 
the environment density. The $R$ values we list are representative of an ISM
medium with a wide range of density values: 
$80\lesssim n_{CSM} \lesssim 8000\,\rm{cm^{-3}}$. }.
Around maximum light (days 8-38), we constrain $n_{CSM}<770 \,\rm{cm^{-3}}$ 
for distances $(1\lesssim R\lesssim3)\times 10^{16}\,\rm{cm}$.
For an ISM scenario our constraints on the particle density $\propto (1/L_{\rm{bol}})^{(1/0.5)}$
(see Appendix \ref{Sec:IClum}). 

Figure \ref{Fig:ICWIND} (lower panel) shows how our \emph{Chandra} 
limit compares to deep radio observations
of SN~2011fe. We explore a wide parameter space to understand how a different
photon effective temperature and/or electron power-law index $p$ would affect 
the inferred density limit: we find $n_{CSM}\lesssim 150 \,\rm{cm^{-3}}$
for $T_{\rm{eff}}<20000$ K and $2.2\lesssim p \lesssim 3$.  X-ray observations 
are less constraining than radio observations in the ISM case when compared
to the wind case: this basically reflects the higher sensitivity of the synchrotron radio
emission to the blastwave velocity, which is faster for an ISM-like ambient 
(for the same density at a given radius).

\subsection{Implications}
\label{SubSec:implications}

From the \emph{Chandra} non detection we derive 
$\dot M/v_w<2\times 10^{-9} (M_{\sun}y^{-1}/100\,\rm{km\,s^{-1}})$. This is the deepest
limit obtained from X-ray observations to date and directly follows from (i) unprecedented
deep \emph{Chandra} observations, (ii) proximity of SN~2011fe coupled to (iii) a consistent
treatment of the dynamics of the SN shock interaction with the environment (Appendix 
\ref{Sec:IClum}). Before SN~2011fe, the deepest X-ray non-detection was reported for 
Type Ia SN~2002bo at a level of $\sim2\times 10^{38}\,\rm{erg\,s^{-1}}$ (distance of 22 Mpc): 
using 20 ks of \emph{Chandra}  observations obtained $9.3\,\rm{days}$ after explosion,
\cite{Hughes07} constrained $\dot M/v_w\lesssim 10^{-4} (M_{\sun}y^{-1}/100\,\rm{km\,s^{-1}})$.
This limit was computed conservatively assuming thermal emission as the
leading radiative mechanism in the X-rays.
Using a less conservative approach, other studies were able to constrain the X-ray 
luminosity from type Ia SNe observed by \emph{Swift} 
to be $\lesssim 10^{39}\,\rm{erg\,s^{-1}}$ \citep{Immler06}, leading to  
$\dot M/v_w\lesssim 10^{-7}(M_{\sun}y^{-1}/100\,\rm{km\,s^{-1}})$  (a factor $\sim100$
above our result). 

Our limit on SN~2011fe strongly argues against a symbiotic binary progenitor for \emph{this}
supernova. According to this scenario the WD accretes material from the wind
of a giant star carrying away material at a level of $\dot M>10^{-8}M_{\sun}yr^{-1}$
for $v_w\lesssim 100\,\rm{km\,s^{-1}}$ (see e.g. \citealt{Seaquist90}; \citealt{Patat11a}; \citealt{Chen11}). 
We reached the same conclusion in our companion paper \citep{Chomiuk12}
starting from deep radio observations of SN~2011fe. The radio limit is
shown in Fig. \ref{Fig:ICWIND} for the range of values $0.01<\epsilon_B<0.1$,
with $\epsilon_B=0.1$ leading to the most constraining limit (where $\epsilon_B$
is the post shock energy density fraction in magnetic fields). 
Historical imaging at the SN site rules out red-giant stars and the majority of the
parameter space associated with He star companions (\citealt{Li11}, their Fig. 2): 
however, pre-explosion images could \emph{not} constrain the Roche-lobe overflow (RLOF)
scenario, where the WD accretes material either from a subgiant or a
main-sequence star. In this case, winds or transferred material lost at the outer Lagrangian
points  of the system are expected to contribute at a level 
$\gtrsim3\times10^{-9} (\dot M/M_{\rm{\sun}}yr^{-1})(v_w/100\rm{km\,\rm{s^{-1}}})^{-1}$
\emph{if} a fraction $\gtrsim1\%$ of the transferred mass is lost at the Lagrangian points
and the WD is steadily burning (see e.g. \citealt{Chomiuk12} et al and references therein). 
The real fraction value is however highly uncertain,
so that it seems premature to rule out the entire class of models based on the
present evidence. X-ray limits would be compatible with RLOF scenarios where
the fraction of lost material is $<1\%$ 
(for any $2.1\lesssim p\lesssim 3$ and $5000\,\rm{K}\lesssim T_{\rm{eff}}\lesssim 20000$ K,
Fig. \ref{Fig:ICWIND}).  However, from the analysis of early UV/optical data, \cite{Bloom11} 
found the companion radius
to be $R_c<0.1 R_{\sun}$, thus excluding Roche-lobe
overflowing red-giants and main sequence secondary stars (see also \citealt{Brown11}).

X-ray non-detections are instead consistent  with (but can hardly be considered a proof of) 
the class of double degenerate (DD) models for type Ia SNe, where two WDs in a close
binary system eventually merge due to the emission of gravitational waves.
No X-ray emission is predicted (apart from the shock break out at $t\ll 1\,\rm{day}$, see Sec. 
\ref{Sec:Gammaray}) 
and SN~2011fe might be embedded in a 
constant and low-density environment (at least for $R>10^{14}\,\rm{cm}$). 
Pre-explosion radio HI imaging indicates an ambient density of $\approx 1\, \rm{cm^{-3}}$
\citep{Chomiuk12} (on scales $R>>10^{14}\,\rm{cm}$), while our tightest limits in the case of an ISM environment  
are $n_{\rm{CSM}}<166\,\rm{cm^{-3}}$.  Our observations cannot however
constrain the presence of material at distances in the range $10^{13}-10^{14}\,\rm{cm}$
from the SN explosion:  recent studies suggest that significant material from the 
secondary (disrupted) WD may indeed reside at those distances either 
as a direct result of the DD-merger \citep{Shen11} or as an outcome of the
subsequent evolution of the system \citep{Fryer10}. 

Whatever the density profile of the environment, our findings are suggestive of a 
\emph{clean} environment around SN~2011fe for distances 
$2\times 10^{15}<R<5\times 10^{16}\,\rm{cm}$. The presence of significant material
at larger distances ($R\gtrsim 5\times10^{16}\,\rm{cm}$) cannot be excluded, so that
our observations cannot constrain models that predict a large delay  ($\gtrsim 10^5$ yr)
between mass loss and the SN explosion
(see e.g. \citealt{Justham11}, \citealt{DiStefano11} and references therein).
Finally, it is interesting to note that the high-resolution spectroscopy study 
by \cite{Patat11b} lead to a similar, \emph{clean} environment conclusion:
at variance with SN~2006X \citep{Patat07}, SN~1999cl \citep{Blondin09} and SN~2007le 
\citep{Simon09}, SN~2011fe shows no evidence for variable sodium absorption
in the time period $8-86$ days since explosion. In this context,
a recent study by \cite{Sternberg11} found evidence for gas outflows from
Type Ia progenitor systems in at least $ 20\%$ of cases. 

Independent constraints on the circumstellar medium density  around Type Ia SNe 
come from Galactic Type Ia  supernova remnants (SNR): 
the study of Tycho's  SNR in the X-rays lead \cite{Katsuda10} to determine a pre-shock 
ambient density of less than $\sim 0.2\,\rm{cm^{-3}}$;  the 
ambient density is likely $<1\,\rm{cm^{-3}}$ both in the case of  Kepler's SNR \citep{Vink08}
and in the case of SNR 0509-67.5 \citep{Kosenko08}.

We emphasize that different type Ia SNe might have different progenitor systems 
as suggested by the increasing evidence of diversity among this class: we know that 
30\% of local SNe Ia have peculiar optical properties (\citealt{Li11b}, \citealt{Li01}). The above
discussion directly addresses the progenitor system of SN~2011fe: our conclusions 
cannot be extended to the entire class of type Ia SNe.
\section{Limits on the post-shock energy density}
\label{Sec:epsilonB}

\begin{figure}
\vskip -0.0 true cm
\centering
    \includegraphics[scale=0.45]{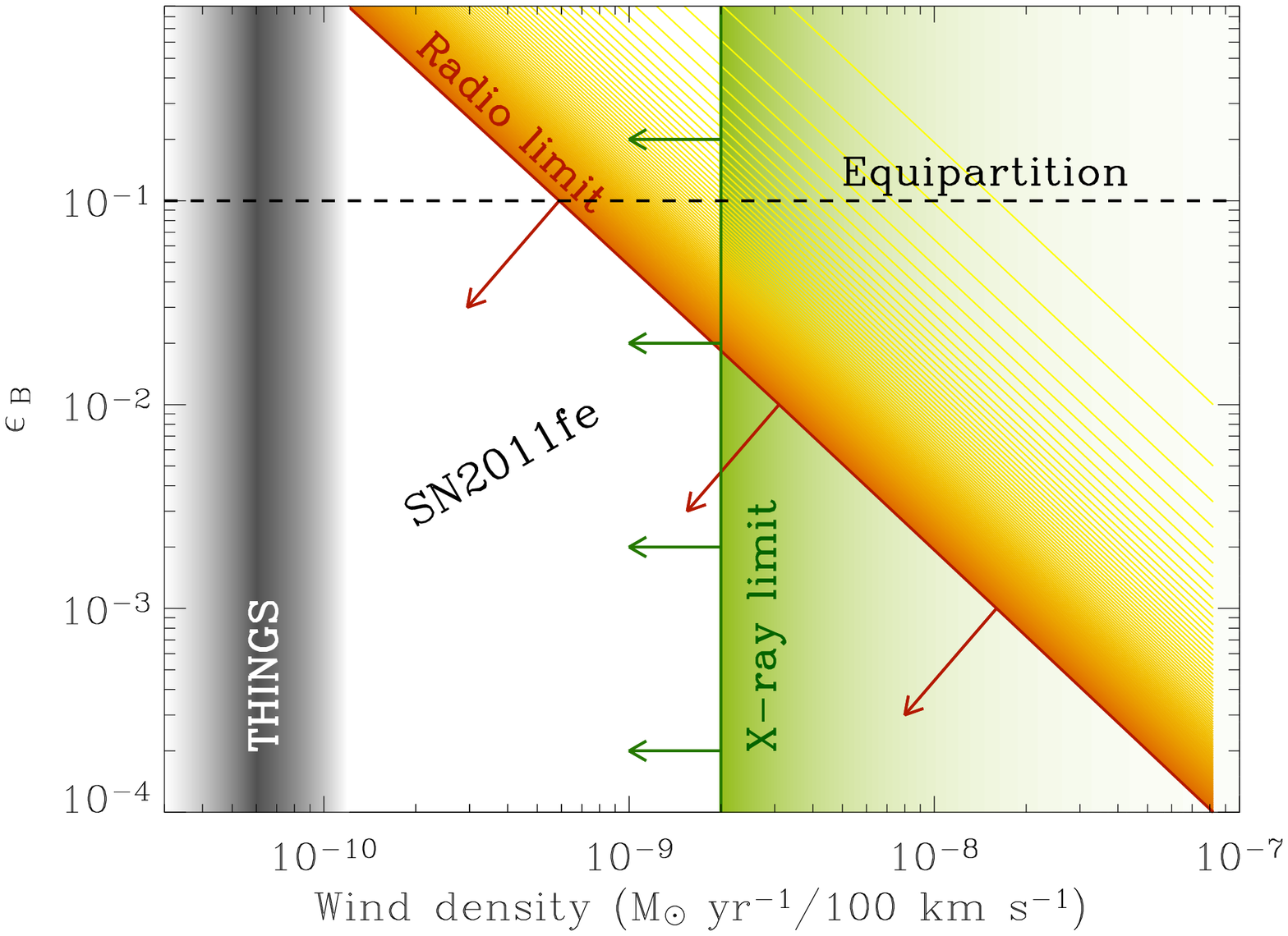} 
    \includegraphics[scale=0.45]{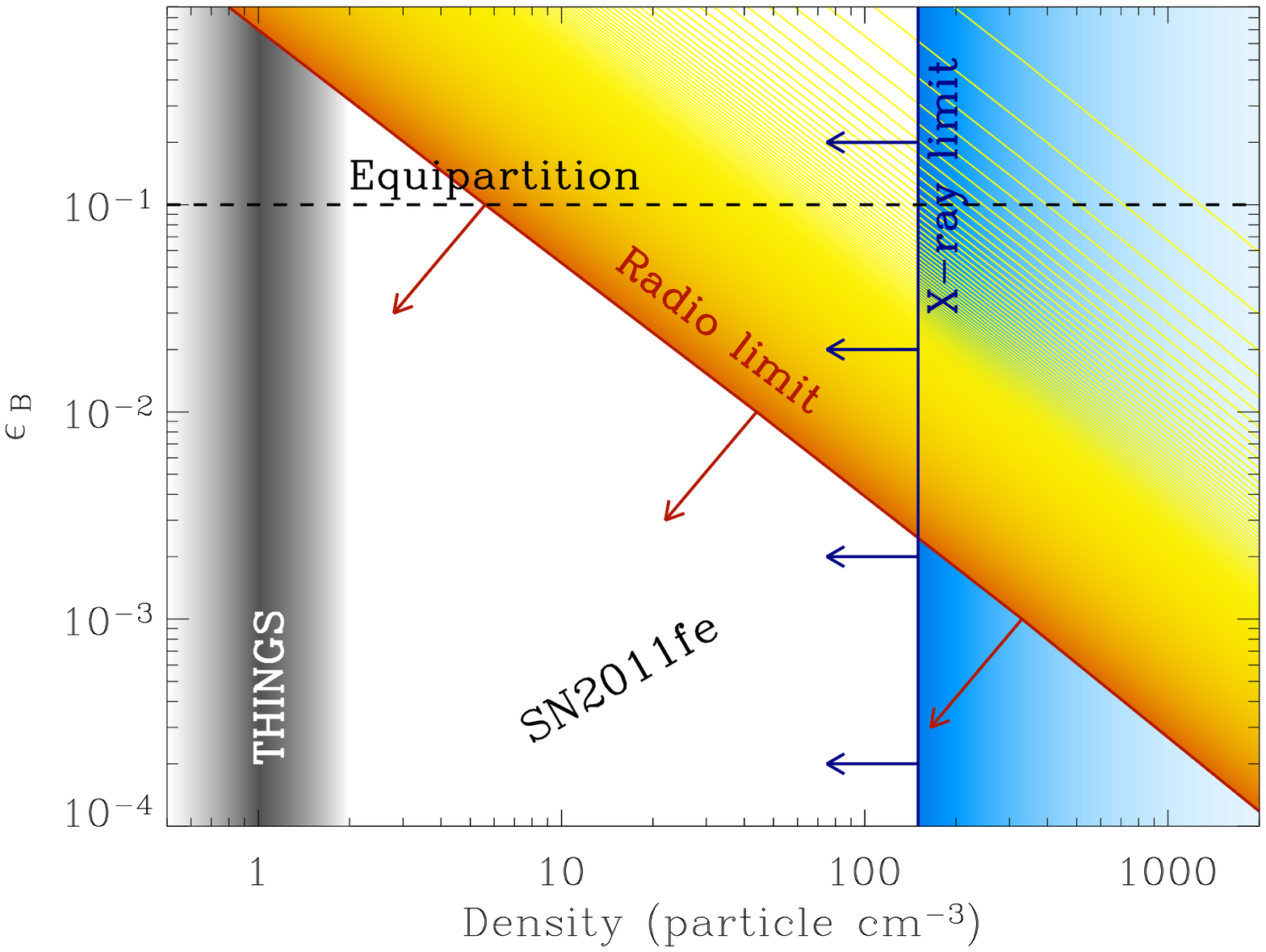}
     \caption{Constraints on the post-shock energy density in magnetic fields vs. ambient density parameter space
     as obtained combining the X-ray to the radio limits from \cite{Chomiuk12}. \emph{Upper panel}: wind
     scenario. \emph{Lower panel}: 
     ISM environment. In both panels the grey area marks the pre-explosion density as measured from 
     radio observations at the SN site \citep{Chomiuk12}. A distance of $4\times 10^{15}\,\rm{cm}$ 
     has been used in the case of a wind medium. The horizontal dashed line marks equipartition
     ($\epsilon_B=\epsilon_e$) for the assumed $\epsilon_e=0.1$. THINGS stands for 
     ``The HI Nearby Galaxy Survey" \citep{Walter08}.}
\label{Fig:epsilonB}
\end{figure}

While the IC emission model discussed here is primarily sensitive to CSM density, the 
associated radio synchrotron emission is sensitive to both the
CSM density and $\epsilon_B$ (post shock energy density in magnetic fields). 
As a consequence, when combined with radio observations of synchrotron self-absorbed SNe,
deep X-ray limits can be used to constrain the
$\epsilon_B$ vs. ambient density parameter space (\citealt{Chevalier06}; \citealt{Katz12}). This is shown in 
Fig. \ref{Fig:epsilonB} for a wind (upper panel) and ISM (lower panel) environment around
SN~2011fe: the use of the same formalism (and assumptions) allows us to directly combine the
radio limits from \cite{Chomiuk12} with our results. We exclude the values of  $\epsilon_B<0.02$ coupled to $\dot M>2\times
10^{-9}\,\rm{M_{\sun}y^{-1}}$ for a wind medium, while $\epsilon_B<0.1$ for any
$\dot M>5\times10^{-10}\,\rm{M_{\sun}y^{-1}}$. In the case of an ISM profile, 
X-ray limits rule out the  $\epsilon_B<2\times10^{-3}$  $n_{CSM}>150\,\rm{cm^{-3}}$
parameter space.

The exact value of the microphysical parameters $\epsilon_B$ and $\epsilon_e$
is highly debated both in the case of non-relativistic (e.g. SNe) 
and relativistic (e.g. Gamma-Ray Bursts, GRBs) shocks: equipartition ($\epsilon_B/\epsilon_e\sim1$) 
was obtained for SN~2002ap from a detailed modeling of the X-ray and radio 
emission \citep{Bjornsson04} while significant departure from equipartition ($\epsilon_e/\epsilon_B\approx 30$) 
has recently been suggested by \cite{Soderberg11} to model SN~2011dh. The same
is true for SN~1993J, for which $\epsilon_B/\epsilon_e\gg 1$ \citep{Fransson98}.
In the context of relativistic shocks, GRB afterglows seem to exhibit a large range
of $\epsilon_B$ and $\epsilon_e$ values (e.g. \citealt{Panaitescu01});  furthermore, values
as low as $\epsilon_B\sim10^{-5}$ have recently  been be suggested by \cite{Kumar10}
from accurate multi-wavelength modeling of GRBs with GeV emission. It is at the moment unclear
if this is to be extended to the entire population of GRBs. 
On purely theoretical grounds, starting from relativistic MHD simulations \cite{Zhang09}
concluded $\epsilon_B\sim5\times10^{-3}$: this result applies to
GRB internal shocks, the late stage of GRB afterglows, transrelativistic SN explosions
(like SN~1998bw, \citealt{Kulkarni98}) and shock breakout from Type Ibc supernova (e.g. SN~2008D, 
\citealt{Soderberg08}).  It is not clear how different  the magnetic field generation and particle acceleration might be
between relativistic and non-relativistic shocks.

Figure \ref{Fig:epsilonB} constitutes the first attempt to infer  the $\epsilon_B$  value
combining deep radio and X-ray observations of a Type Ia SN:  
better constraints on the parameters could in principle be obtained \emph{if} X-ray observations
are acquired at the SN optical maximum light. In the case of SN~2011fe we estimate that 
a factor $\sim10$ improvement on the density limits would have been obtained with 
a \emph{Chandra} observation at maximum light.
\section{Gamma and X-ray emission from shock break out}
\label{Sec:Gammaray}

Shock break out from WD explosions is expected to produce a short 
($\approx 1-30\,\rm{ms}$) pulse with typical $\sim\,\rm{MeV}$ photon energy,  
luminosity $\sim10^{44}\,\rm{erg\,s^{-1}}$ and energy in the range 
$10^{40}-10^{42}\,\rm{erg}$ \citep{Nakar11}.  Such an emission episode would
be easily detected if it were to happen close by (either in the Milky Way or in the 
Magellanic Clouds), while SN~2011fe exploded $\sim6.4$ Mpc away \citep{Shappee11}.
Given the exceptional proximity of SN~2011fe we nevertheless
searched for evidence of high-energy emission from the shock break-out
using data collected by the nine spacecrafts of the interplanetary network (IPN
Mars Odyssey, Konus-Wind, RHESSI, INTEGRAL (SPI-ACS),
\emph{Swift}-BAT, Suzaku, AGILE, MESSENGER, and Fermi-GBM).  

The IPN is full sky with temporal duty cycle $\sim100\%$ and is sensitive to radiation in the
range $20-10^4$ keV \citep{Hurley10}.
Within a 2-day window centered on Aug 23rd a total of 3 bursts were detected
and localized by multiple instruments of the IPN. 
Out of these 3 confirmed bursts, one has localization consistent with 
SN~2011fe. Interestingly, this burst was detected by KONUS, Suzaku and INTEGRAL (SPI-ACS) 
on August 23rd 13:28:25 UT: for comparison, the inferred explosion time of SN~2011fe is
$16:29\pm20$ minutes, \citealt{Nugent11}. The IPN error box area for this burst is
$1.4$ sr.  The poor localization of this event does not allow us to firmly associate 
this burst with SN~2011fe: from poissonian statistics we calculate a $\sim10\%$
chance probability for this burst to be spatially consistent with SN~2011fe.
A more detailed analysis reveals that SN~2011fe lies inside the KONUS-INTEGRAL
triangulation annulus but outside the KONUS-Suzaku triangulation annulus.
Furthermore, at the inferred time of explosion, SN~2011fe was slightly above the 
Fermi-GBM horizon, but no burst was detected (in spite of the stable GBM
background around this time). We therefore conclude that there is no statistically significant
evidence for a SN-associated burst  down to the Fermi-GBM threshold
(fluence $\sim4\times 10^{-8}\,\rm{erg\,cm^{-2}}$ in the 8-1000 keV band)\footnote{\emph{Swift} 
is sensitive to fainter bursts: however it has a limited temporal coverage. 
We note that \emph{Swift}-BAT was active and no burst was detected during the time window 
extending from 16:03:54 UT  to 16:30:53 UT, implying a probability $>50\%$ for a SN-associated
burst with fluence above the \emph{Swift} threshold and below the Fermi-GBM one to occur 
without being detected.}.

The early photometry of SN~2011fe constrains the progenitor radius to be $R_p\lesssim 0.02\,R_{\sun}$ 
\citep{Bloom11}. Using the fiducial values $E=10^{51}\,\rm{erg}$, $M_{ej}=1.4\,M_{\sun}$, the shock 
break out associated with SN~2011fe is therefore expected to have released
$E_{\rm{BO}}\lesssim3\times10^{41}\,\rm{erg}$ over a time-scale 
$t_{\rm{BO}}\lesssim2\,\rm{ms}$ with luminosity 
$L_{\rm{BO}}\gtrsim7\times10^{43}\,\rm{erg\,s^{-1}}$ at typical 
$T_{\rm{BO}}\gtrsim250\,\rm{keV}$ (see \citealt{Nakar11}, their Eq. 29).
At the distance of SN~2011fe, the expected fluence is as low as 
$\sim5\times10^{-11}\,\rm{erg\,cm^{-2}}$ which is below the threshold of 
all gamma-ray observatories currently on orbit (the weakest burst observed
by BAT had a 15-150 keV fluence of $\sim6\times10^{-9}\,\rm{erg\,cm^{-2}}$).
For comparison, the KONUS-Suzaku-INTEGRAL burst formally consistent with the position of 
SN~2011fe was detected with fluence $\sim3\times10^{-6}\,\rm{erg\,cm^{-2}}$
and duration of a few seconds (peak flux of $\sim 4\times10^{-7}\,\rm{erg\,s^{-1}cm^{-2}}$). 
If it were to be connected with the SN, the associated 
$3-$sec peak luminosity would be $L\sim2\times10^{45}\rm{erg\,s^{-1}}$ and total energy 
$E\sim10^{46}\rm{erg}$ (quantities computed in the 20-1400 keV energy band)
which are orders of magnitudes above expectations.

For $t>t_{\rm{BO}}$, the temperature and luminosity drop quickly (see \citealt{Nakar11}
for details): in particular, for $t>t_{\rm{NW}}$ the emitting shell enters the Newtonian phase.
For SN~2011fe we estimate $t_{\rm{NW}}\sim0.3\rm{s}$ (\citealt{Nakar11}, their Eq. 30); for $R_p\lesssim 0.02\,R_{\sun}$ 
the luminosity at  $t=10\times t_{\rm{NW}}$ is  $L(t_{\rm{NW}})\gtrsim 1\times10^{41}\,\rm{erg\,s^{-1}}$
with typical emission in the soft X-rays: $T(t_{\rm{NW}})\gtrsim0.2\,\rm{keV}$.
At later times $L\propto t^{-0.35}$ \citep{Nakar11} while $T$ rapidly drops  below the
\emph{Swift}-XRT energy band (0.3-10 keV).  \emph{Swift}-XRT observations
were unfortunately not acquired early enough to constrain the shock break out 
emission from SN~2011fe. 
UV observations were not acquired early enough either: after $\sim1$ hr
the UV emission connected with the shock break out is expected to be
strongly suppressed due to the deviation from pure radiation domination (e.g. \citealt{Rabinak11}).  
It is however interesting to note the presence of a "shoulder" in the UV light-curve 
\citep{Margutti11b} particularly prominent in the uvm2 filter for $t<4$ days (see
\citealt{Brown11}, their Fig. 2) whose origin is still unclear (see however
\citealt{Piro12}). A detailed modeling is required
to disentangle the contribution of different physical processes to the early UV emission
(and understand which is the role of the "red leak" -see e.g. \cite{Milne10}- 
of the uvm2 filter in shaping the observed light-curve).

The collision of the SN ejecta with the companion star is also expected to produce X-ray emission
with typical release of energy $E_{x}\sim 10^{46}-10^{47}\,\rm{erg}$ in the hours following 
the explosion (a mechanism which has been referred to as the analog of shock break out 
emission in core collapse SNe, \citealt{Kasen10}). According to \cite{Kasen10}, 
in the most favorable scenario of a red-giant companion of $M\sim 1\,M_{\sun}$ at separation 
distance $a=2\times 10^{13}\,\rm{cm}$, the interaction time-scale is $\sim5\,\rm{hr}$ after the SN explosion
and the burst of X-ray radiation lasts $1.9\,\rm{hr}$ 
(with a typical luminosity $\sim6\times10^{44}\,\rm{erg\,s^{-1}}$): too short to be caught
by our \emph{Swift}-XRT re-pointing 1.25 days after the explosion. We  furthermore
estimate the high energy tail of the  longer lasting thermal optical/UV emission associated 
to the collision with the companion star to be too faint to be detected either: at
$t\sim1.5\,\rm{days}$, the emission has $T_{\rm{eff}}\lesssim25000\,\rm{K}$ 
and peaks at frequency $\nu \lesssim 3\times 10^{15}\,\rm{Hz}$ (Eq. 25 from \citealt{Kasen10}). 
Non-thermal particle acceleration might be a source of X-rays at these times,
a scenario for which we still lack clear predictions:
future studies will help understand the role of non-thermal emission in the case
of the collision of a SN with its companion star.
\section{Conclusion}
\label{Sec:conc}

IC emission provides solid limits to the 
environment density which are \emph{not} dependent  on assumptions about the poorly 
constrained magnetic field energy density (i. e. the $\epsilon_B$
parameter; see also \citealt{Chevalier06} and \citealt{Horesh11}).
This is different from the synchrotron emission, which was used in our companion paper
\citep{Chomiuk12} to constrain the environment of the same event from
the deepest radio observations ever obtained for a SN Ia. 
The two perspectives are complementary:
the use of the same assumptions and of a consistent formalism furthermore allows
us to constrain the post-shock energy density in magnetic fields vs. ambient density
parameter space (see Fig. \ref{Fig:epsilonB}). 
This plot shows how deep and contemporaneous radio and 
X-rays observations of SNe might be used to infer  the shock parameters. 

The IC luminosity is however strongly dependent on the SN bolometric luminosity:
$L_{\rm{IC}}(t)\propto L_{\rm{bol}}(t)$. Here we presented the deepest limit on the
ambient density around a type Ia SN obtained from X-ray observations.  Our results
directly benefit from: (i) unprecedented
deep \emph{Chandra} observations of one of the nearest type Ia SNe, coupled to (ii) a consistent
treatment of the dynamics of the SN shock interaction with the environment (Appendix 
\ref{Sec:IClum} and \citealt{Chomiuk12}), together with (iii) the direct computation of the SN bolometric 
luminosity from \emph{Swift}/UVOT data.

In particular we showed that:
\begin{itemize}
\item Assuming a wind profile the X-ray non-detections imply a mass loss 
$\dot M<2\times10^{-9}\,\rm{M_{\sun}yr^{-1}}$ for $v_w=100\,\rm{km\,s^{-1}}$. This 
is  a factor of $\sim 10$ deeper than the limit reported by
\citealt{Horesh11}. This rules out symbiotic binary progenitors for SN~2011fe and argues against  Roche-lobe
overflowing  subgiants and main sequence secondary stars \emph{if} a fraction
$\gtrsim1\%$ of the transferred mass is lost at the Lagrangian points and the WD is
steadily burning.
\item Were SN~2011fe to be embedded in an ISM environment, our calculations 
constrain the density to $n_{CSM}<160\,\rm{cm^{-3}}$.
\end{itemize}

Whatever the density profile, the X-ray non-detections are suggestive of
a \emph{ clean} environment around SN~2011fe, for distances in the range
$\sim (0.2-5)\times 10^{16}\,\rm{cm}$.  This is either consistent with 
the bulk of material (transferred from the donor star to the accreting WD or resulting
from the merging of the two WDs) to be confined within 
the binary system or with a significant delay $\gtrsim 10^5$ yr between mass loss 
and SN explosion (e.g.  \citealt{Justham11}, \citealt{DiStefano11}). Note that in the 
context of DD mergers, the
presence of material on distances $10^{13}-10^{14}\,\rm{cm}$ (as recently 
suggested by e.g. \citealt{Fryer10} and \citealt{Shen11}) has been excluded
by \cite{Nugent11} based on the lack of bright, early UV/optical emission.

We furthermore looked for bursts of gamma-rays associated with the shock break out
from SN~2011fe. We find no statistically significant evidence for a SN-associated burst for fluences
$>6\times10^{-7}\,\rm{erg\,cm^{-2}}$. However, with progenitor radius $R_p<0.02~R_{\sun}$
the expected SN~2011fe shock break out fluence is 
$\approx5\times10^{-11}\,\rm{erg\,cm^{-2}}$, below the sensitivity of gamma-ray detectors
currently on orbit. 

The proximity of SN~2011fe coupled to the sensitivity of \emph{Chandra} observations, 
make the limits presented in this paper difficult to be surpassed in the near future for type Ia SNe. 
However, the generalized IC formalism of Appendix \ref{Sec:IClum} is 
applicable to the entire class of hydrogen poor SNe, and will provide the tightest constraints
to the explosion environment \emph{if} X-ray observations are acquired around maximum light 
(see Fig. \ref{Fig:IC}) for Type I supernovae (Ia, Ib and Ic).

\acknowledgments
We thank Harvey Tananbaum and Neil Gehrels for making \emph{Chandra} and
\emph{Swift} observations possible.
We thank Re'em Sari, Bob Kirshner, Sayan Chakraborti, Stephan Immler, Brosk Russel and 
Rodolfo Barniol Duran for helpful discussions.
L.C. is a Jansky Fellow of the National Radio Astronomy Observatory.
R.J.F. is supported by a Clay Fellowship.
KH is grateful for IPN support under the following NASA grants:
NNX10AR12G (Suzaku), NNX12AD68G (Swift), NNX07AR71G (MESSENGER),
and NNX10AU34G (Fermi). The Konus-Wind experiment is supported by a Russian 
Space Agency contract and RFBR grant 11-02-12082-ofi\_m. 
POS acknowledges partial support from NASA Contract NAS8-03060.
\bibliographystyle{fapj}

\appendix
\section{Inverse Compton luminosity}
\label{Sec:IClum}
Ambient electrons accelerated to relativistic speed by the SN shock are expected
to upscatter optical photons from the SN photosphere to X-ray frequencies via Inverse
Compton (IC), see e.g. \cite{Chevalier06b}, \cite{Chevalier06}. Here we generalize Eq.
(31) from \cite{Chevalier06} for a population of relativistic electrons with arbitrary
distribution $n_e(\gamma)=n_0\gamma^{-p}$ for $\gamma>\gamma_{\rm{min}}$, 
both for an ISM (Eq. \ref{Eq:ICISM}) and a wind (Eq. \ref{Eq:ICWIND}) scenario.

Using the IC emissivity given by \cite{Felten66}, their Eq. 27, the IC luminosity reads:
\begin{equation}
\label{Eq:one}
	\frac{dL_{\rm{IC}}}{d\nu}=2.1\sigma_{\rm{T}} c \Big ( \frac{h}{3.6k}\Big )^{\frac{3-p}{2}}R^2 n_0 \Delta R \rho_{\rm{rad}} 
	T_{\rm{eff}}^{\frac{p-3}{2}} \nu^{\frac{1-p}{2}}
\end{equation}
where $\rho_{\rm{rad}}(t)=\frac{L_{\rm{bol}}(t)}{4\pi R^2 c}$ is the energy density of photons of 
effective temperature $T_{\rm{eff}}$ which are upscattered to 
$\sim3.6\gamma^2 kT_{\rm{eff}}$; $\Delta R$ is the extension of the region containing
fast electrons while $R$ is the (forward) shock radius. The emission is expected to originate from 
a shell of shocked gas between the reverse and the forward shock which are separated by 
the contact discontinuity at $R_{c}$ \citep{Chevalier06}. For $\rho_{\rm{SN}}\propto R^{-n}$ with
$n=10$ the forward shock is at $1.239R_{c}$ ($1.131R_{c}$) while the reverse shock is at $0.984R_c$ 
($0.966R_c$) in the case of a wind (ISM) environment \citep{Chevalier82}. The fraction of 
the volume within the forward shock with shocked gas is $0.5$  ($0.4$)  
corresponding to a sphere of radius $\Delta R\sim 0.8 R$ ($\Delta R\sim 0.7 R$)
for an assumed wind (ISM) density profile.

If a fraction $\epsilon_e$ of the post-shock energy density goes into non thermal
relativistic electrons, from 
$\int_{\gamma_{\rm{min}}}^{\infty}\gamma\cdot n_e(\gamma)d\gamma= 9/8 \epsilon_e \rho_{CSM}v_s^2$
we have:
\begin{equation}
\label{Eq:two}
	n_0=\frac{9(p-2) \epsilon_e \rho_{CSM} v_s^2 \gamma_{\rm{min}}^{(p-2)}}{8m_e c^2}
\end{equation}
for $p>2$. Combining Eq. \ref{Eq:one} with Eq. \ref{Eq:two}, we obtain Eq.  \ref{Eq:ICgeneral}. The temporal evolution
of $L_{\rm{IC}}$ directly depends on $L_{\rm{bol}}(t)$; $T_{\rm{eff}}(t)$; $v_s(t)$; $R(t)$ and $\gamma_{\rm{min}}(t)$.
The properties of the SN and of its progenitor determine $L_{\rm{bol}}(t)$, $T_{\rm{eff}}(t)$ and the profile of the outer ejecta
 $\rho_{\rm{SN}}\propto R^{-n}$. We assume $n\sim 10$ through out the paper
(e.g. \citealt{Chevalier06}). The environment sets the $\rho_{CSM}$ profile, which we parametrize as  $\rho_{CSM}\equiv A\cdot R^{-s}$.
Both the SN explosion properties \emph{and} the environment determine the shock dynamics: evolution of the shock radius
$R(t)$, shock velocity $v_s(t)$ and, as a consequence $\gamma_{\rm{min}} (t)$. Under those conditions the shock
interaction region can be described by a self-similar solution \citep{Chevalier82} with the shock radius evolving as
$R\propto t^{(\frac{n-3}{n-s})}$ which implies:
\begin{equation}
	\label{Eq:vshockRshock}
	v_s(t)=\Big ( \frac{n-3}{n-s}\Big ) \frac{R(t)}{t}	
\end{equation}
The shock velocity directly determines $\gamma_{\rm{min}}$. From \cite{Soderberg05},
assuming that \emph{all} electrons go into a power-law spectrum with spectral index $p$:
\begin{equation}
	\label{Eq:gammamin}
	\gamma_{\rm{min}}(t)= \frac{9\epsilon_e}{8\eta} \Big ( \frac{m_p}{m_e}\Big ) \Big ( \frac{v_s(t)}{c}\Big )^2
	 \Big ( \frac{\mu_i}{N_e/N_i}\Big ) \Big ( \frac{p-2}{p-1}\Big )
\end{equation}
where $\eta$ is the shock compression parameter, $N_e$ ($N_i$) is the electron (ion) number density and $\mu_i$
is the average number of nucleons per atom.
We furthermore define $g(Z)\equiv \Big ( \frac{\mu_i}{N_e/N_i}\Big )$. For Solar metallicity $g(Z_{\sun})\approx1.22$.
In the following we assume $\eta\approx 4$ \citep{Chevalier06}, $Z=Z_{\sun}$. 
\subsection{ISM scenario:}
The self-similar solutions for the interaction of the SN ejecta with an ISM-like circumstellar medium ($s=0$, $\rho_{CSM}\equiv A/R^s=A$) lead to (\citealt{Chevalier82}, Soderberg et al., in prep):
\begin{equation}
	\label{Eq:vshockISM}
	v_s(t)=2.4\times 10^9 \Big (\frac{A}{\rm{g\,/cm^3}}  \Big)^{-0.1} \Big ( \frac{E}{10^{51}\,\rm{erg}} \Big)^{0.35}
	\Big(\frac{M_{ej}}{1.4 M_{\sun}} \Big)^{-0.25}\Big(\frac{t}{s} \Big)^{-0.29}\,\,\, \rm{cm\,s^{-1}}
\end{equation}
where $M_{ej}$ is the mass of the ejected material and $E$ is the energy of the supernova explosion. 
Eq. \ref{Eq:two}, \ref{Eq:vshockRshock}, 
\ref{Eq:gammamin} and \ref{Eq:vshockISM}, together with Eq. \ref{Eq:one}, predict an IC luminosity:
\begin{equation}
	\label{Eq:ICISM}
	\frac{dL_{\rm{IC}}}{d\nu}= f_{\rm{ISM}}(p, Z) \epsilon_e^{p-1}
	\Big (\frac{M_{ej}}{1.4M_{\sun}} \Big)^{\frac{1-2p}{4}}
	\Big (\frac{A}{\rm{g\,cm^{-3}}} \Big)^{(1.1-0.2p)}
	\Big (\frac{E}{10^{51}\rm{erg}} \Big)^{(0.7p-0.35)}
	\Big (\frac{t}{s} \Big)^{(1.29-0.58p)}
	T_{\rm{eff}}^{\frac{p-3}{2}}\nu^{\frac{1-p}{2}}
	 \Big (\frac{L_{\rm{bol}}}{\rm{erg\,s^{-1}}}\Big ) \,\,\rm{\frac{erg}{s\,Hz}}
\end{equation}
with $f_{\rm{ISM}}(p,Z)\approx2.0\times 10^7(10^3)^{(1.1-0.2p)}(1.3\times10^{-11})^{\frac{3-p}{2}}\Big (\frac{53.9}{2+p}\Big)^{(p-2)}(p-2)^{(p-1)}g(Z)^{(p-2)}$.
In the body of the paper $A$ will be reported in (hydrogen) particles per $\rm{cm^{3}}$.
\subsection{WIND scenario:}
For $s=2$ ($\rho_{CSM}\equiv A/R^2$) the self-similar solutions lead to (\citealt{Chevalier82}, Soderberg et al., in prep):
\begin{equation}
	\label{Eq:vshockWIND}
	v_s(t)=6.6\times10^{11} \Big( \frac{A}{\rm{g/cm}}\Big)^{-0.12}\Big( \frac{E}{10^{51}\rm{erg}}\Big)^{0.43}\Big( \frac{M}{1.4M_{\sun}}\Big)^{-0.31}\Big( \frac{t}{s}\Big)^{-0.12}\,\,\, \rm{cm\,s^{-1}}
\end{equation}
Combining Eq. \ref{Eq:two}, \ref{Eq:vshockRshock}, \ref{Eq:gammamin} and \ref{Eq:vshockWIND} with Eq. \ref{Eq:one} we obtain:
\begin{equation}
	\label{Eq:ICWIND}
	\frac{dL_{\rm{IC}}}{d\nu}= f_{\rm{WIND}}(p, Z) \epsilon_e^{p-1}
	\Big (\frac{M_{ej}}{1.4M_{\sun}} \Big)^{(0.93-0.62p)}
	\Big (\frac{A}{\rm{g\,cm^{-1}}} \Big)^{(1.36-0.24p)}
	\Big (\frac{E}{10^{51}\rm{erg}} \Big)^{(0.86p-1.29)}
	\Big (\frac{t}{s} \Big)^{-(0.24p+0.64)}
	T_{\rm{eff}}^{\frac{p-3}{2}}\nu^{\frac{1-p}{2}}
	\Big (\frac{L_{\rm{bol}}}{\rm{erg\,s^{-1}}}\Big ) \,\,\rm{\frac{erg}{s\,Hz}}
\end{equation}
with $f_{\rm{ISM}}(p, Z)\approx6.7\times10^{-7}10^{(0.24p-1.36)}(1.3\times10^{-11})^{\frac{3-p}{2}}\Big (\frac{5.6\times10^5}{2+p}\Big)^{(p-2)}(p-2)^{(p-1)}g(Z)^{(p-2)}$.\\
Note that $\rho_{CSM}\equiv A/R^2\equiv \dot M/ (4\pi v_w R^2)$, so that $A=\dot M/(4\pi v_w)$, where $\dot M$ and $v_w$ are the mass loss rate and the 
wind velocity of the SN progenitor, respectively. In the body of the paper, for the wind scenario, we refer to $A$ in terms of mass loss rate for a given wind velocity so that 
it is easier to connect our results to known physical systems.
\end{document}